\documentclass[]{aa}

\usepackage{psfig}

\begin{document}


\title{Effect of the environment on the Fundamental Plane  of
elliptical galaxies} 

\author{E.A.~Evstigneeva\inst{1}, V.P.~Reshetnikov\inst{1,2}, 
N.Ya.~Sotnikova\inst{1}}

\offprints{E.A.~Evstigneeva \hfill\break(e-mail: katya@gong.astro.spbu.ru)}   

\institute{Astronomical Institute of St.Petersburg State
University, 198504 St.Petersburg, Russia
\and
           Isaac Netwon Institute of Chile, St.Petersburg Branch 
}  

\date{Received 11 May 2001 / Accepted 17 September 2001}

\titlerunning{Fundamental Plane}

\abstract{
We present an analysis of the location of interacting E/S0 galaxies on the
Fundamental Plane (FP). Using the NEMO package, we performed N-body
simulations of close encounters and mergers
of two self-gravitating spherical galaxies. Two models for encounting
galaxies -- Plummer's and Hernquist's -- were used. The number of particles
ranged from 20000 to 50000 per galaxy. The changes of central density,
half-mass radius and central velocity dispersion were analysed. It was
found that close encounters between galaxies alter noticeably the above
parameters within a very short time interval (10$^7$--10$^8$ years) just
before the final merger. The amplitudes of parameter changes strongly depend
on the initial mass concentration of a model. In some experiments we considered
an encounter of two spherical galaxies with dark matter components. The
effect of parameter changes was less pronounced than for the experiments
without a dark halo. The results of the simulations were used to discuss the
FP for interacting early-type galaxies.
}

\authorrunning{E.Evstigneeva, V.Reshetnikov, N.Sotnikova}
 
\maketitle  

\section{Introduction}

The Fundamental Plane (FP) defines one of the most important relationships
for early-type galaxies (Djorgovski \& Davis 1987, Dressler et al. 1987). 
The FP combines surface photometry parameters
($R_e$ -- effective radius and $\mu_e$ -- effective surface brightness
or $\langle \mu \rangle_e$ -- mean surface brightness within $R_e$)
with spectroscopy characteristics (line-of-sight central velocity
dispersion $\sigma_0$). The measured values of $R_e$, $\mu_e$ and
$\sigma_0$ for a sample of E and S0 galaxies do not fill this
3-parameter space entirely but rather a thin plane within it
(with scatter $\sim$ 0.1 dex). The FP can be projected onto any two
axes. Examples of these projections are the Kormendy relationship
($\mu_e$--log$R_e$), and the Faber-Jackson relationship between
luminosity and velocity dispersion.

The interpretation of the FP is still a matter of discussion.
The most popular point of view is that the FP is simply a
consequence of the Virial theorem and the fact that E/S0 galaxies
have similar mass-to-luminosity ratios ($M/L$) and homologous structure
at a given luminosity. If elliptical galaxies had perfectly homologous 
structural and kinematical properties, then the Virial theorem would predict 
the FP to have the form $R_e \propto \sigma_0^2 \Sigma_e^{-1} (M/L)^{-1}$,
where $\Sigma_e = 10^{-0.4 \langle \mu \rangle_e}$.
Under the assumption of a constant $M/L$, the observed FP should then be 
$\rm{log} R_e \propto 2 \rm{log} \sigma_0 + 0.4 \langle \mu \rangle_e$. 
In practice, however, the $\rm{log} \sigma_0$
and $\mu_e$ coefficients have been determined to lie in the range 1.2-1.6, 
and 0.30-0.35, respectively. The difference between expected and 
observed parameters of the FP is generally refered to as the
"tilt" of the FP. This tilt 
is usually viewed as the product of systematic variations of $M/L$ along the FP
($M/L \propto L^{\beta}$). Among the possible explanations that have been 
considered for the tilt are: variations in the stellar population along 
the FP, changes in the dark matter content, and deviations from homology
(Hjorth \& Madsen 1995, Capelato et al. 1995, Ciotti et al. 1996).

The very existence of the FP has profound implications for the processes that 
led to the formation and evolution of elliptical galaxies. In particular, its 
tightness provides a very strong constraint on the evolutionary history of 
those systems. The FP can be used to study the evolution of galaxies as a 
function of redshift, to derive redshift independent distance estimates, etc.

However, we must be cautious when using the FP to 
obtain distances to elliptical galaxies without regard for their environment.
The significant environmental effects on the elliptical galaxy
correlations prevent their utility as accurate distance indicators.
Hence the question of how a galaxy should move in the space of 
($R_e$, $\mu_e$, $\sigma_0$) as a result of interaction with other galaxies
is of a great importance.

Zepf \& Whitmore (1992) investigated the environmental 
dependence of the FP of elliptical galaxies by studying a sample of elliptical 
galaxies in {\it compact groups}. The central velocity dispersions 
of ellipticals in compact groups appear to be about 20 \% lower 
than those in other environments (galaxies in clusters and in the general field). 
De la Rosa et al.(2001) re-examined this topic with a better and more 
homogeneous dataset. They found little or no significant 
difference between elliptical galaxies in compact groups and comparable 
galaxies in other environments.

Levine \& Aguilar (1996) carried out a few Monte Carlo simulations of the 
evolution of a number of elliptical galaxies in the core of a very rich 
{\it cluster}. 
Under two very simple physical assumptions (virial equilibrium and retention of 
the functional form of the surface brightness and mass density profiles), any 
change in the galaxy caused by any interaction that satisfies these 
assumptions will be narrowly constrained within a plane very close to the 
observed FP of ellipticals, even for large changes in the global $M/L$ ratio. 
The angle between this plane and the observed FP of ellipticals is less than 
$15^{\rm o}$. So, if a galaxy is in or near the FP, it will remain 
so after an interaction.

De Carvalho \& Djorgovski (1992) found that elliptical galaxies in clusters 
were best fitted by a slightly different FP than galaxies in the field.
However according to Pahre et al. (1998) this difference is naturally 
explained by errors in estimating dust extinction or estimating distances 
(and not by any intrinsic differences in the stellar populations among 
elliptical galaxies that correlates with their environment).
There is no {\it clear} evidence that either the slope, zero-point
or scatter of the FP are dependent on the spatial environment (see
discussion in Pahre et al. 1998).

The main purpose of the present work is to explore the possible influence of
{\it strong mutual low-velocity encounters} of early-type galaxies on 
their general characteristics and on their location within the FP.

\section{Fundamental Plane for Interacting and Merging Galaxies}

In order to study the FP for intensively interacting early-type
galaxies we have considered several pieces of observational data.

Fig.1 presents the Faber-Jackson relation for E/S0 galaxies
belonging to 1) VV (Vorontsov-Velyaminov 1959) or Arp (Arp 1966) 
systems (28 galaxies, stars),
2) CPG (Karachentsev 1972) or TURNER binary systems (Turner 1976), and 
3) Hickson compact groups (Hickson 1982) 
(circles). The total number of galaxies is 94.

\begin{figure}
\centerline{\psfig{file=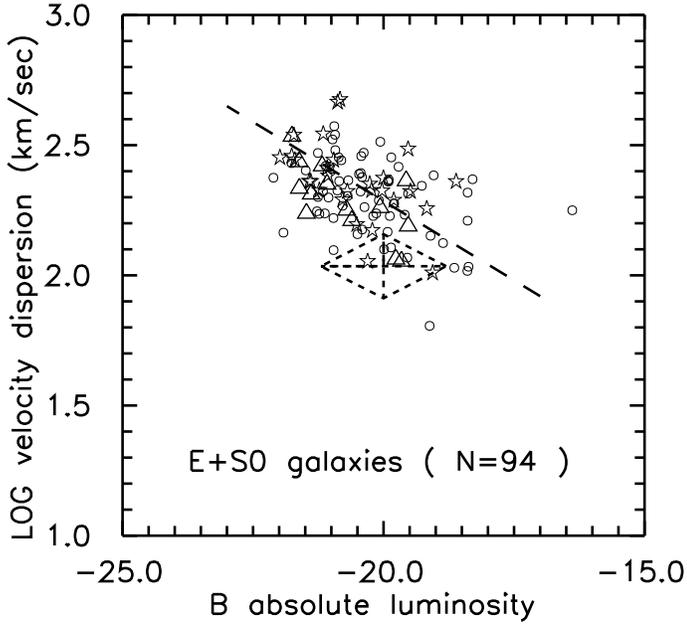,width=10cm,clip=}}
\caption{The Faber-Jackson relation for 94 interacting E/S0 galaxies.}
\end{figure}

We used $\sigma_0$ from 
1)~Hypercat, CRAL Observatoire de Lyon (www-obs.univlyon1.fr/hypercat),
2)~Davoust \& Considere (1995), 
3)~Simien \& Prugniel (1997a),
4)~Simien \& Prugniel (1997b)
and apparent magnitudes $m_B$ from NED\footnote{NASA/IPAC Extragalactic 
Database (NED) operated under the management of the Jet Propulsion 
Laboratory, CalTech, in accordance with a contract from NASA}.
The $m_B$ have been corrected for galactic extinction using NED 
(reddenings derived from HI and galaxy counts: Burstein \& Heiles 1982). 
Absolute magnitudes M$_B$ are computed assuming $H_0$ = 75 km s$^{-1}$ Mpc$^{-1}$. 
When estimating galaxy distances, 
we assigned to each galaxy the heliocentric radial velocity adopted from NED 
and corrected for the Sun's motion using the standard relation. 

Dashed line in Fig.1 shows the mean relation for 594 E/S0 galaxies
according to McElroy (1995): $L_B \propto \sigma_0^{3.3}$.
The slope and the scatter ($\sigma(\rm{log} \sigma_0)=0.15$)
of the data for interacting galaxies are the same as for normal
E/S0.

Triangles in Fig.1 present the characteristics of 14 remnants of
disk-disk mergers (Keel \& Wu 1995).
As one can see, the mergers follow the same relation as early-type
galaxies but with small shift ($\sim 0.^m6$) to brighter absolute
luminosities. The dashed rhomb shows the mean value ($\pm 1 \sigma$)
for 9 starbursting infrared-luminous galaxies (Shier \& Fischer 1998). 
All these galaxies are at some stage of merging.
Starbursting mergers demonstrate significant brightening
($\sim 2^m$) in comparison with normal galaxies. One can expect
that relics of disk-disk mergers and infrared-luminous mergers
will evolve into normal ellipticals at the FP.

Fig.2 compares the FP location of interacting ellipticals/lenticulars belonging to
VV and Arp systems (crosses) with relatively isolated galaxies
(points) (the data in the $r_G$ passband are from Djorgovski \& Davies 1987, 
in the $B$ filter -- Bender et al. 1992). Here we used $H_0$ = 100 km s$^{-1}$ Mpc$^{-1}$. 
The FP is defined as 
$\rm{log} R_e \propto \rm{log} \sigma_0 + 0.26 \langle \mu \rangle_e$.
The FP scatter for interacting galaxies 
($\sigma (\rm{log} R_e) = 0.09$ in $B$ and $\sigma (\rm{log} R_e) = 0.14$ in $r_G$) 
is the same as (or even smaller than) that of non-interacting objects
($\sigma (\rm{log} R_e) = 0.19$ in $B$ and $\sigma (\rm{log} R_e) = 0.15$ in $r_G$).

\begin{figure*}
\centerline{\psfig{file=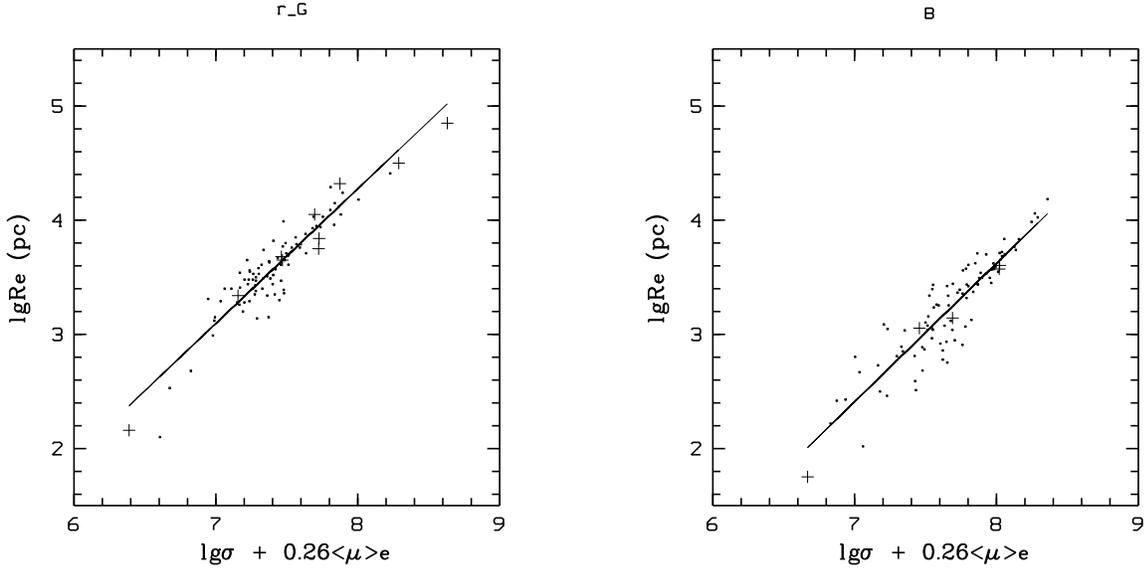,width=16cm,angle=-90,clip=}}
\caption{The FP for interacting E/S0 galaxies.}
\end{figure*}

One can conclude that the present observational data about the
global characteristics of strongly interacting E/S0 galaxies show
that the close environment has no or almost negligible effect on the
FP. The only difference we see for forming (or young) ellipticals 
is that they are significantly brighter (for fixed $\sigma_0$)
due to superimposed bursts of star formation.

\section{Numerical Modeling of Interacting Ellipticals}

To check our empirical findings we have performed
numerical modeling of close encounters of two ellipticals.

We simulated the evolution of the self-gravitating spherical galaxies by
using the NEMO package (http://bima.astro.umd/nemo/, Teuben 1995). This is a
freeware package designed to numerically solve gravitational N-body
problems. It consists of subroutines for specifying initial configurations 
of stellar-dynamical systems (including many standard models) and
subroutines for simulating the evolution of these systems based on various
numerical schemes. 

In our computations, we used N-body dynamics
with a hierarchical tree algorithm and multipole expansion to compute the
forces, as proposed by Barnes \& Hut (1986). Multipole
expansion is limited by the monopole term. A tolerance parameter $\theta$
("opening angle") which controlled the force computation from distant
particles was set as 0.7. 

We considered only the behavior of the stellar component of galaxies (elliptical 
galaxies usually contain a small amount of gas).
The number of particles used in our numerical simulations ranged from
$N_{tot} = 20\,000$ to $N_{tot} = 50\,000$ per galaxy. In this case we managed to
substantially suppress the effects of pair relaxation and to trace the
evolution of merging galaxies on time scales up to
$t \approx 0.5\times10^9$ years. The timestep for the simulations used was 
as $\Delta t = 10^6$ years.

The softening length $\epsilon$ was taken as $< 0.1$ of the inter-particle 
distance of the homogoneous state.

We considered two models for encounting galaxies. One of them is Plummer's
spherically-symmetric model

\begin{eqnarray}
\Phi(r) &=&
- \frac{G\, M}{(r^2+a_{pl}^2)^{1/2}} \, ,
\nonumber\\
&&
\nonumber\\
\rho(r) &=&
\frac{3 M}{4 \pi} \, \frac{a_{pl}^2}{\left(r^2 + a_{pl}^2\right)^{5/2}} \, ,
\nonumber
\end{eqnarray}
where $M$ is the galaxy mass and $a_{pl}$ is a scale length.

In some experiments galaxies are modeled by the potential-density pair proposed
by Hernquist (1990) for spherical galaxies

\begin{eqnarray}
\Phi(r) &=&
- \frac{G\, M}{r + a_{hq}} \, ,
\nonumber\\
&&
\nonumber\\
\rho(r) &=&
\frac{M}{2 \pi a_{hq}^3} \, \frac{a_{hq}^4}{r\,(r+a_{hq})^3}\, .
\nonumber
\end{eqnarray}

It is well known that both models are described by a distribution function
in an analytical form and, in the absence of numerical errors and dynamical
instabilities, remain time-stationary.

\begin{figure*}[!ht]
\centerline{\psfig{file=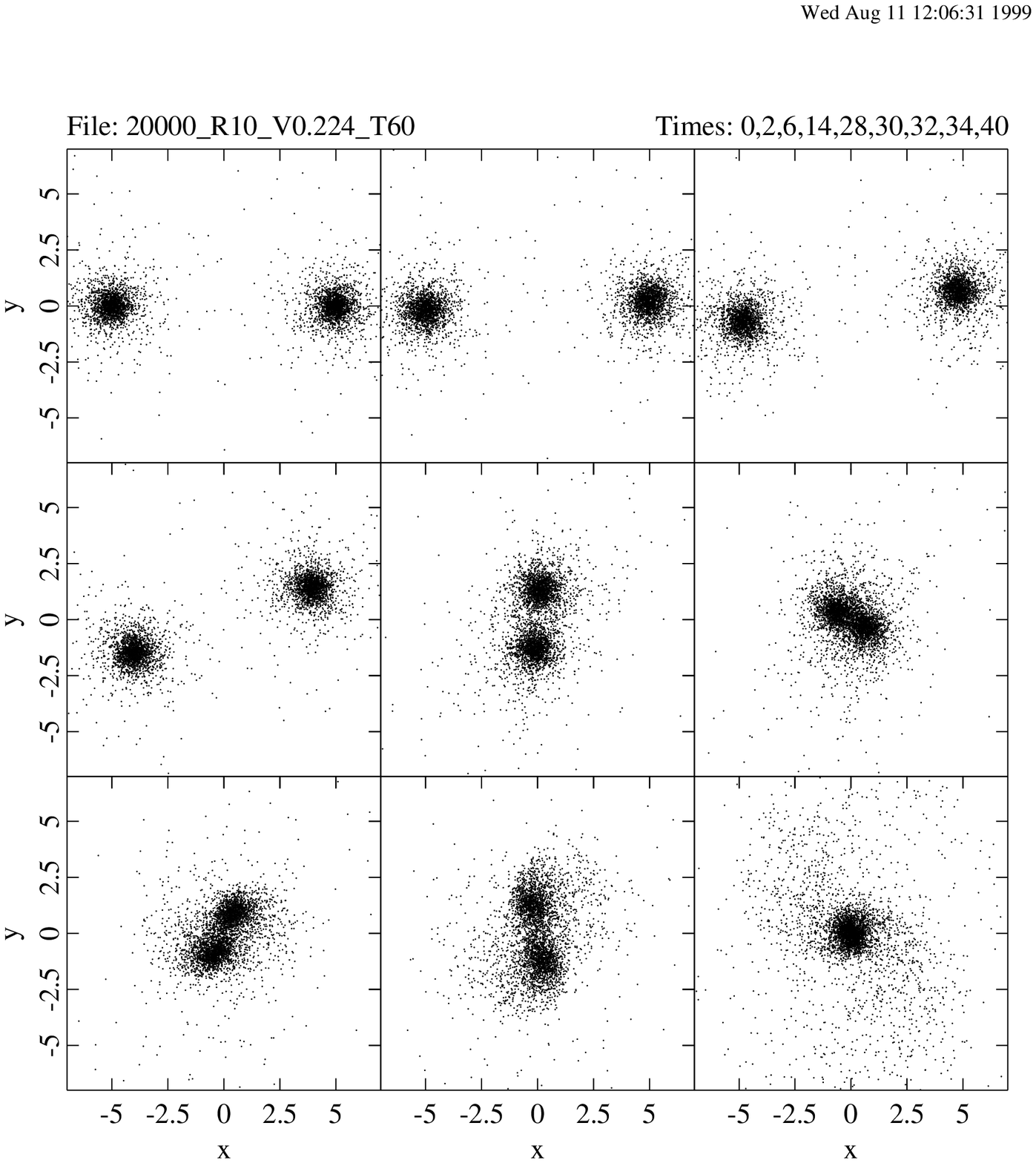,width=14cm,clip=}}
\caption{Close encounter of two identical Plummer's spheres. 
Times: 0,2,6,14,28,30,32,34,40 (left to right, top to bottom).}
\end{figure*}

\begin{figure*}[!ht]
\centerline{\psfig{file=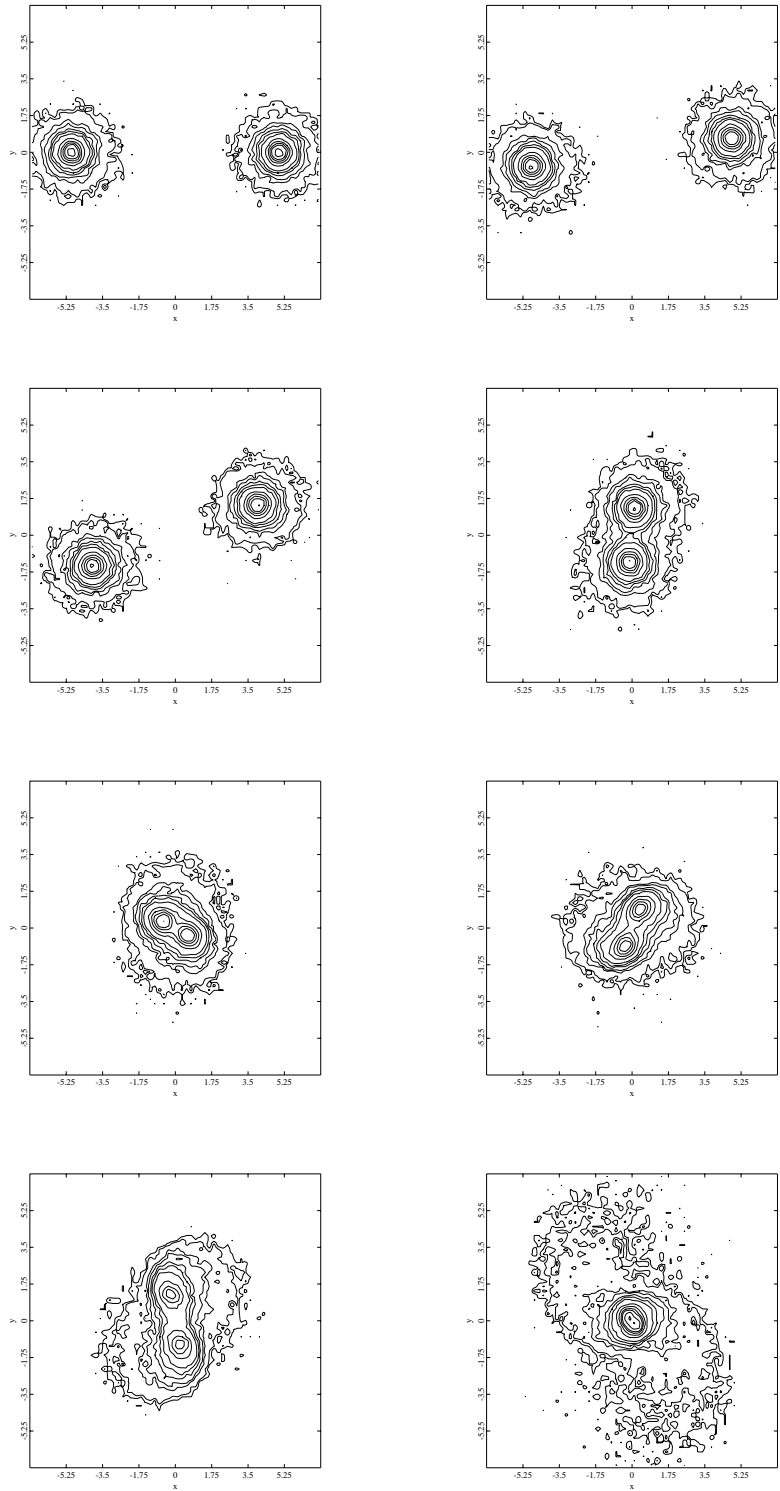,height=22cm,clip=}}
\caption{Contour maps of two interacting Plummer's spheres. 
Times: 0,6,14,28,30,32,34,40 (left to right, top to bottom).}
\end{figure*}

\begin{figure*}[!ht]
\centerline{\psfig{file=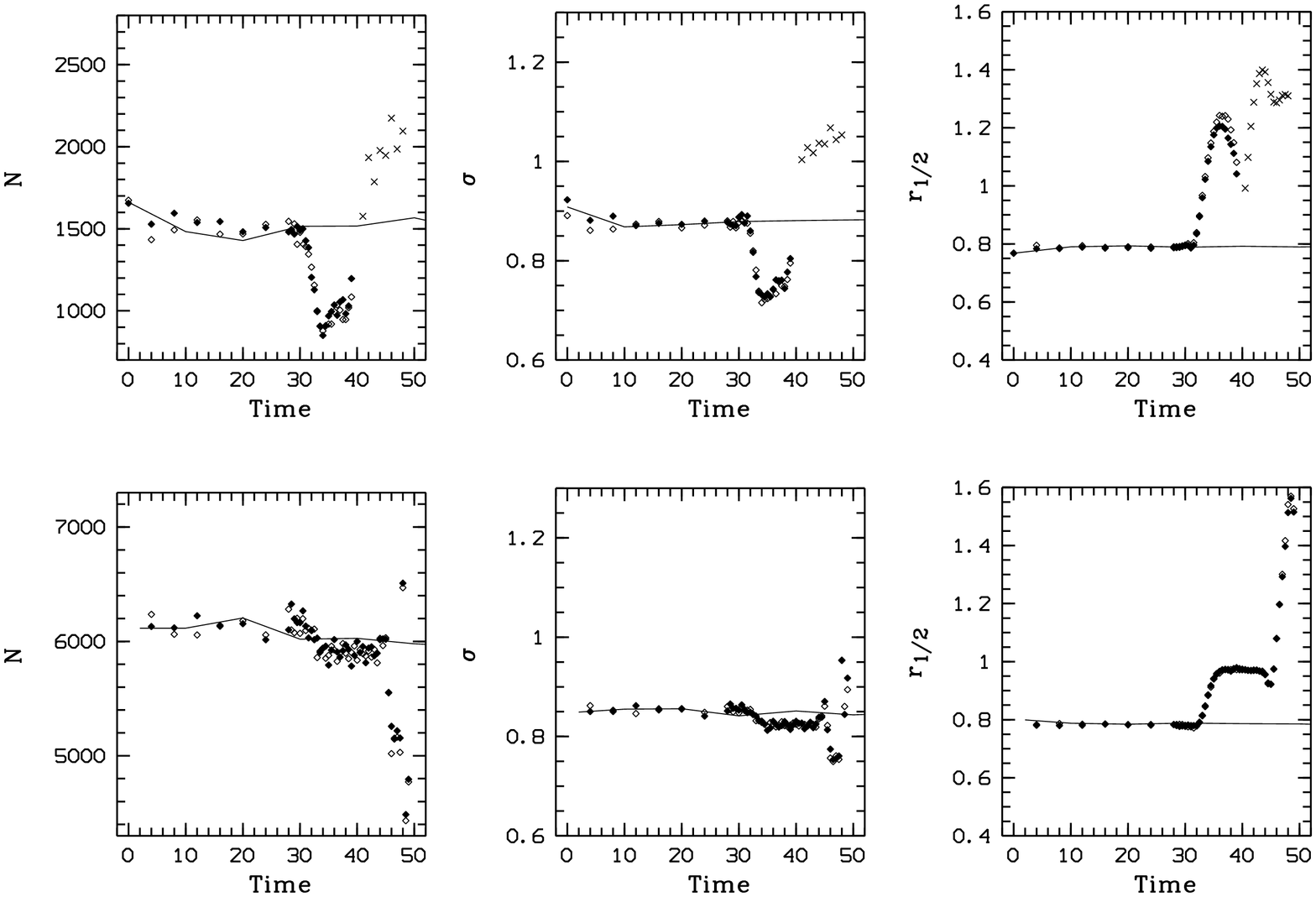,angle=-90,width=16cm,clip=}}
\caption{Evolution of the characteristics of two model galaxies
(solid and open rhombs) during a close encounter ($N_{tot}$=50\,000 per
galaxy). Top three figures -- Plummer's spheres, bottom -- Hernquist's spheres.
Crosses show parameters after merging of two Plummer's spheres in one object.
Evolution of the corresponding isolated spheres are shown by solid lines.
$t=0$ corresponds to initial configuration, $t$=30,32 -- first
pericenter passage, $t$=34,38 -- the time of largest separation after
the pericenter passage and $t\approx$40,49 -- merging (first and second
numbers refer to Plummer's and Hernquist's models correspondingly).}
\end{figure*}

Results are presented in the following system of units: gravitational
constant $G = 1$, the galaxy mass $M = 1$, the total energy of a sphere
$E= -1/4$ ($E= - 3 \pi G M^2/ 64 a_{pl}$ and $a_{pl} = 3 \pi / 16$ for Plummer's sphere; 
$E= - G M^2/ 12 a_{hq}$ and $a_{hq} = 1/3$ for Hernquist's sphere). Scaled to 
physical values appropriate for a typical
elliptical galaxy, i.e. $M = 10^{11}$ and half-mass radius
$r_{1/2} = 3$ kpc ($r_{1/2}\approx1.31 a_{pl}$ - for Plummer's sphere,
$r_{1/2}\approx2.41 a_{hq}$ - for Hernquist's sphere), units of distance, time
and velocity are 3.73 kpc, 10.3 Myr, 345.3 km s$^{-1}$ respectively.

We specified the initial distance between two equal-mass
galaxies as $R = 37.3$ kpc and chose
the initial relative velocity in the range $V = 77.3-103.6$ km s$^{-1}$.

As a result we had a close encounter with merging (the distance of the
first closest approach was 5.2 kpc) and a distant encounter without merging
(in this case the minimum galaxy separation was 10.3 kpc). 

Fig.3 presents some ''snapshots'' of a close encounter, showing the initial 
condition ($t=0$), the configuration near the first maximum overlap ($t=30$), the
configuration shortly before the final merger ($t=34$), and a merger state
($t=40$). A contour map of two interacting Plummer's spheres are plotted in Fig.4. 
Some changes in morphology of model galaxies become noticeable only at the final 
stages of encounter ($t=30-34$). There are real objects which are very
similar to these interaction stages of model galaxies, for example 
NGC~1587/1588, NGC~7236/7237 (Borne \& Hoessel 1988).

For all models we followed the changes of central density ($N$ -- number of 
particles within 0.2 distance units = 0.75 kpc), half-mass radius 
($r_{1/2}$ -- within 0.75 kpc) and central velocity dispersion ($\sigma_0$). 
During the distant encounter all these
parameters were not altered. As to the close encounter, in this case the
most drastic changes in parameters were observed at the moment shortly
before the final merger (at $\Delta\,t \approx 5 \approx 50\,\times\,10^6$\,\,yrs 
before merger) -- Fig.5. 

The range of parameter changes appears to depend on the initial mass 
concentration of a model. More concentrated Hernquist's spheres 
show less pronounced responses than does a less concentrated Plummer's model (Fig.5). 

Amplitudes of relative changes of the parameters for two models are:\\
Plummer's sphere --
$\delta r_{1/2}$ $\approx$ 67$\%$, $\delta \sigma_0$ $\approx$ 17$\%$, \\
Hernquist's sphere --
$\delta r_{1/2}$ $\approx$ 33$\%$, $\delta \sigma_0$ $\approx$ 6$\%$.\\
These amplitudes are comparable with the FP scatter (rms scatter
$\sim$0.1 -- 0.15 dex). Therefore even the close low-velocity
encounters cannot significantly enlarge dispersion of the observed FP.
Moreover the galaxies demonstrate large changes only within a very short time
interval ($\sim$10$^7$--10$^8$ years) just before the final merger 
(short relative to the evolution time of merging galaxies).

It should be noted that Hernquist's sphere shows large parameter deviations 
from initial values (more than the values given above) at the very last 
stages of the encounter before merger ($t = 45 - 49$) -- Fig.5. We do not discuss 
these moments because of the short duration of these periods and 
the difficulty in distinguishing almost merged galaxies. 

We have also considered an encounter between two galaxies embedded in dark
halos. Each galaxy was composed of a spherical component, represented by a
Plummer's model, and a spherical Plummer's halo. The mass of dark matter was set 
equal to the visible mass within the sphere containing 95 \% of the stellar mass.
The combined potential of both components was used to solve the Jeans equations 
and to derive the initial velocity dispersion (Hernquist 1993).  
Each sphere consisted of $N_{tot} = 40\,000$ particles ($N_{tot} = 20\,000$ per stellar 
component and $N_{tot} = 20\,000$ per halo). 

The addition of an extended dark halo surrounding each galaxy makes the
effect of parameter changes less pronounced than for the experiments without
a dark halo. Amplitudes of relative changes of the parameters are:\\
Plummer's sphere --
$\delta r_{1/2}$ $\approx$ 60$\%$, $\delta \sigma_0$ $\approx$ 25$\%$, \\
Plummers's sphere with halo --
$\delta r_{1/2}$ $\approx$ 30$\%$, $\delta \sigma_0$ $\approx$ 15$\%$.\\

\section{Conclusions}

Our numerical experiments and observational data have shown that the global 
parameters of early-type galaxies are rather stable to gravitational 
perturbation. The FP parameters do change during close encounters but within a very 
short time interval ($\sim$10$^7$--10$^8$ years) just before the final merger. 
Furthermore, the amplitudes of these changes are comparable to the scatter 
of the observed FP of ellipticals. There is a very small probability that we 
observe the system at these interaction stages.
Therefore, the FP is still a good method to derive distances, even to the 
clusters that include a large number of interacting galaxies
(e.g. van Dokkum et al. 1999).  

\acknowledgements{We acknowledge support from the "Integration"
programme (A0145). We would like to thank the referee, Dr. R. de Carvalho,
for useful comments and suggestions.}


\begin{thebibliography}{}

\bibitem{} Arp, H. 1966, Atlas of Peculiar Galaxies, Pasadena
\bibitem{} Barnes, J. \& Hut, P. 1986, Nature, 324, 446
\bibitem{} Bender, R., Burstein, D. \& Faber, S.M. 1992, ApJ, 399, 462
\bibitem{} Borne, K.D. \& Hoessel, J.D. 1988, ApJ, 330, 51
\bibitem{} Burstein, D. \& Heiles, C. 1982, AJ, 87, 1165
\bibitem{} Capelato, H.V., de Carvalho, R.R. \& Carlberg, R.G. 1995, ApJ, 451, 525
\bibitem{} Ciotti, L., Lanzoni, B. \& Renzini, A. 1996, MNRAS, 282, 1
\bibitem{} Davoust, E. \& Considere, S. 1995, A\&AS, 110, 19
\bibitem{} de Carvalho, R.R. \& Djorgovski, S. 1992, ApJ, 389, L49
\bibitem{} de la Rosa, I.G., de Carvalho, R.R. \& Zepf, S.E. 2001, AJ, in press
(astro-ph/0104324)
\bibitem{} Djorgovski, S. \& Davis, M. 1987, ApJ, 313, 59
\bibitem{} Dressler, A., Linden-Bell, D., Burstein, D., et al. 1987, 
ApJ, 313, 42
\bibitem{} Hernquist, L. 1990, ApJ, 356, 359
\bibitem{} Hernquist, L. 1993, ApJ, 86, 389
\bibitem{} Hickson, P. 1982, ApJ, 255, 382
\bibitem{} Hjorth, J. \& Madsen, J. 1995, ApJ, 445, 55
\bibitem{} Karachentsev, I.D. 1972, Catalogue of isolated pairs of galaxies in 
the northern hemisphere, Soobshch. Spets. Astrof. Obs 7, 1
\bibitem{} Keel, W.C. \& Wu, W. 1995, AJ, 110, 129
\bibitem{} Levine, S.E. \& Aguilar, L.A. 1996, MNRAS, 280, 13
\bibitem{} McElroy, D.B. 1995, ApJS, 100, 105
\bibitem{} Pahre, M.A., de Carvalho, R.R. \& Djorgovski, S.G. 1998, 
AJ, 116, 1606
\bibitem{} Shier, L.M. \& Fischer, J. 1998, ApJ, 497, 163
\bibitem{} Simien, F. \& Prugniel, Ph. 1997a, A\&AS, 122, 521
\bibitem{} Simien F. \& Prugniel, Ph. 1997b, A\&AS 1126, 15
\bibitem{} Teuben, P. 1995, ASP Conf. Ser., 77, 398
\bibitem{} Turner, E.L. 1976, ApJ, 208, 20
\bibitem{} van Dokkum, P.G., Franx, M., Fabricant, D. et al. 1999, ApJ, 520,
L95
\bibitem{} Vorontsov-Velyaminov, B.A. 1959, Atlas and catalogue of interacting 
galaxies, Sternberg Institute, Moscow State University
\bibitem{} Zepf, S.E. \& Whitmore, B.C. 1993, ApJ, 418, 72
\end{thebibliography}
\end{document}